\documentclass[letterpaper, 11pt]{article}
\pdfoutput = 1

\usepackage{shortcuts}

\usepackage[margin = 2.5cm]{geometry}
\begin{document}

\thispagestyle{empty}

\begin{center}

\vspace*{5em}

{\LARGE \bf Chiral and Geometric Anomalies in Finite Systems}

\vspace{1cm}

{\large \DJ or\dj e Radi\v cevi\'c}
\vspace{1em}

{\it Perimeter Institute for Theoretical Physics, Waterloo, Ontario, Canada N2L 2Y5}\\
\texttt{djordje@pitp.ca}\\

\vspace{0.08\textheight}
\begin{abstract}
Dirac fermions coupled to gauge fields can exhibit the chiral anomaly even on a finite spatial lattice. A careful description of this phenomenon yields new insights into the nature of spin-charge relations and on-site symmetries (symmetries that are gauged by placing gauge fields on all links of the lattice). One notable result is that only sufficiently small symmetry groups can act on-site in a system with finitely many degrees of freedom. Symmetries that break this rule either cannot be gauged on an arbitrary lattice, or the gauging decouples the matter fields from gauge fields. These ``anomalies'' are not quantum in nature, and they are diagnosed geometrically, by the volume of the spatial manifold. The familiar particle number U(1) exhibits this kind of anomalous behavior in any finite fermion theory. The chiral anomaly in such a finite system instead manifests itself most simply after gauging the $\Z_2$ fermion parity.

\end{abstract}
\end{center}

\newpage

\section{Introduction and summary}

A symmetry of a quantum theory that cannot be gauged is typically said to be anomalous. Anomalies appear often in  modern physics, playing important r$\hat{\trm{o}}$les in explaining some observed processes in particle physics, constraining spaces of possible theories, classifying phases of matter, and providing consistency checks of conjectured dualities and renormalization group flows \cite{Weinberg:1996kr, Harvey:2005it}.

The chiral or Adler-Bell-Jackiw anomaly is a familiar example of this phenomenon in quantum field theory. It is found in theories of quantum electrodynamics in even spacetime dimensions. Free Dirac fermions in these dimensions have separately conserved currents of opposite chiralities. When a U(1) gauge field is coupled to the fermion theory, quantum effects prevent the difference of these two currents, the axial current, from being conserved \cite{Adler:1969, Bell:1969, Fujikawa:1979ay, Ambjorn:1983hp}. In other words, Gauss operators corresponding to U(1)$\times$U(1)$\^A$ particle and axial number symmetry groups do not commute. The chiral symmetry U(1)$\^A$ then has a quantum anomaly in a theory with a gauged U(1) (see also \cite{Cho:2014jfa, Kapustin:2014zva}).

In all common manifestations of the chiral anomaly, the underlying system is infinite in some sense, and the anomaly is a one-loop effect in some perturbative framework. One goal of this short note is to show how a chiral anomaly can be precisely seen in manifestly \emph{finite} theories.\footnote{This note expressly does \emph{not} attempt to avoid any no-go theorems about putting chiral theories on a lattice.} The setup is completely nonperturbative. No classical limits or low-energy effective field theories are used. The focus is on $d = 1$ spatial dimension, but the results generalize to arbitrary odd $d$.

For concreteness, the discussion will mostly focus on finite-lattice models of spinless complex fer\-mi\-ons, paired up (``dimerized'') into Dirac spinors as per the standard formulation of staggered fermions \cite{Kogut:1974ag, Susskind:1976jm}. In this setup, the axial number is \emph{not} associated with a symmetry that acts on-site, i.e.~the generator of the chiral symmetry is not a product of operators that act on each lattice site separately. A finite system with an anomalous symmetry that is ``off-site'' is not a novelty, as similar  examples were constructed and studied in \cite{Wen:2013oza} and references therein. However, efforts to define the U(1)$\times$U(1)$\^A$ symmetry and its anomaly in a finite theory will reveal new insights.

The most interesting of these new results is a proof, based on the global consistency of possible gauge constraints, that \emph{any \emph{U(1)} on-site symmetry of fermions on a finite lattice must be geometrically anomalous or quasi-anomalous}. (A symmetry is defined to be quasi-anomalous if gauging it projects the matter sector to a \emph{single} state; these anomalies are called geometric because they are diagnosed by the volume of the spatial manifold, i.e.~the number of lattice sites.) Whether the U(1) is geometrically anomalous or ``merely'' quasi-anomalous on a given lattice depends on the charges assigned to the fermion states. Either outcome is avoided if the symmetry is taken to be sufficiently off-site. For a lattice of $N$ sites, the maximal, practically nonanomalous, particle number symmetry group is $\Z_{N + 1}$, taken to be maximally off-site. Any larger group, including U(1), will be geometrically (quasi-)anomalous for some charges, no matter how off-site it acts. This is a \emph{classical} kind of anomaly, and it is distinct from the chiral anomaly, which is quantum in nature.

More generally, the lesson is that lattice systems with a $D$-dimensional Hilbert space on each site face a geometry-dependent obstruction to being coupled to Abelian gauge theories whose gauge groups have order greater than $D$. This underappreciated fact of life must be taken into account in order to understand the story of the chiral anomaly in a finite theory of fermions. Indeed, the most elementary always-definable setup with a chiral anomaly will involve no U(1) gauge fields --- rather, it will involve gauging the fermion parity $(-1)^F$ by adding appropriate $\Z_2$ gauge fields.

The above conclusions do not invalidate existing lattice calculations, in which it is often possible to focus on a subclass of lattices that, for \emph{some} matter charges, do not engender any anomalies. Another coping mechanism is to add additional degrees of freedom in the UV, making the starting lattice effectively infinite. (Another alternative, making the gauge group noncompact, will not be entertained here.) On an infinite lattice, fermions can be grouped into (possibly finitely many) blocks of infinitely many sites. The fermion number in each block can be regarded as the charge of a state in a coarse-grained theory where each block looks like one lattice site. This coarsening effectively gives the kinds of lattice or continuum theories that are typically discussed when fermions are coupled to U(1) gauge fields. From the point of this effective theory, however, infinitely many UV degrees of freedom had to be added to remove the geometric (quasi-)anomaly of the U(1).

Another novelty in this note concerns the concept of a \emph{spin-charge relation} \cite{Wen:2013ue, Seiberg:2016gmd, Seiberg:2016rsg}. Conventionally, this is the statement that states with odd fermion number have odd U(1) charge. On a finite lattice, due to the lack of sensible U(1) symmetries, this relation can be interpreted as saying that the axial and particle number parities coincide. Coupling a gauge field to fermions changes this relation to one in which local particle and axial parities differ, thereby manifesting one sense in which a lattice gauge field has ``spin.'' Moreover, $\Z_2$ gauge fields coupled to fermions are known to play the role of \emph{spin structures} \cite{Radicevic:2018okd}, and so this basic $\Z_2$ chiral anomaly will also be identifiable as a lattice avatar of the gravitational anomaly of chiral theories in two spacetime dimensions \cite{AlvarezGaume:1983ig, Bardeen:1984pm}.

The structure of this paper is straightforward. Section \ref{subsec defs} introduces all the needed definitions of fermion operators and discusses the staggered fermion theory of Dirac spinors. The material is quite standard and aims to provide several concrete examples that should be kept in mind during the more general discussion later on. Some possibly new comments regarding spin-charge relations are at the end of this section. Section \ref{subsec curious} proves that on-site U(1) symmetries must be geometrically \hbox{(quasi-)anomalous} on finite lattices. The argument is very general and applies to arbitrary Hamiltonians with conserved particle numbers.\footnote{The ``off-site rule'' is also formulated here: possible geometric anomalies in fermion number symmetries can be removed if the appropriate symmetry group $\Z_{K + 1}$ is \emph{defined} to act not on-site, but on $K$ sites. This way, for a lattice of any size, the fermion number symmetry with group $\Z_{K + 1}$ either does not exist, or is anomaly-free.} Section \ref{subsec QEDs} goes back to the issue of chiral anomalies in theories of Dirac spinors coupled to various gauge fields, giving criteria for when a chiral anomaly will appear in geometrically nonanomalous theories. The connection to coarse-graining, thermodynamic limits, spin-charge relations, and spin structures and gravitational anomalies is discussed in the final section \ref{sec remarks}.

\newpage


\section{Preliminary definitions} \label{subsec defs}

Consider $N$ sites arranged in a circle, with spinless fermion operators $\psi_v$, $\psi\+_v$ on each site. It will be convenient to fix the expressions involving these operators in terms of elementary Majorana operators and their bilinears. Let
\bel{\label{def psi}
  \psi_v \equiv \frac 12 \left(\chi_v + \i \chi_v'\right).
}
The Majoranas $\chi_v, \chi_v'$ satisfy $\chi^2_v = (\chi_v')^2 = \1$ and anticommute with each other. This can be written as
\bel{
  \{\chi_v, \chi_u\} = \{\chi'_v, \chi'_u\} = 2\delta_{vu}, \quad \{\chi_v, \chi'_u\} = 0.
}
It is useful to work with Majorana bilinears
\bel{\label{def Z S}
  Z_v \equiv \i \chi'_v \chi_v, \quad S_{vu} \equiv -\i \chi'_v \chi_u.
}
Note that $Z_v$ measures the fermion number (or its parity) at site $v$, while $S_{vu}$ enacts fermionic hopping between $u$ and $v$. Their algebraic properties include
\bel{
  Z_v Z_u = Z_u Z_v, \quad Z_v S_{uw} = (-1)^{\delta_{vu} + \delta_{vw}} S_{uw}Z_v, \quad S_{vu} S_{wz} = (-1)^{\delta_{vw} + \delta_{uz}} S_{wz} S_{vu}.
}
The choice of $\psi_v$ in \eqref{def psi} is consistent with taking $\chi_v$ to dualize (via the Jordan-Wigner map) to the spin chain operator $Z_1 \cdots Z_{v - 1} X_v$. Then $Z_v$ is the usual Pauli matrix equal to $(-1)^{n_v}$, where $n_v \equiv \psi\+_v\psi_v$ is the fermion number operator with eigenvalues $0$ and $1$. The complex fermions therefore satisfy
\gathl{
  \{\psi_v, \psi_u\} = \{\psi\+_v, \psi_u\+\} = 0, \quad \{\psi_v, \psi\+_u\} = \delta_{vu},\\
  Z_v = \1 - 2 \psi_v\+ \psi_v = (-1)^{\psi_v\+ \psi_v} \equiv (-1)^{n_v}, \quad S_{vu} = \left(\psi_v\+ - \psi_v\right) \left(\psi_u\+ + \psi_u\right).
}

When $N$ is even, there is a natural but noncanonical way to write any $d = 1$ theory of spinless complex fermions in terms of Dirac (two-component complex) spinors: simply pair up sites $v = 2x - 1$ and $v = 2x$ into a dimer labeled by $x$, and define
\bel{
  \Psi_x^\alpha \equiv \bcol{\psi_{2x - 1}}{\psi_{2x}},\quad x = 1, \ldots, \frac N 2.
}
It is very convenient to define the \emph{chiral generators} of the operator algebra by
\bel{\label{def chiral basis}
  \Psi_x^\pm \equiv \frac1 {\sqrt 2} \left( \Psi_x^1 \pm \Psi_x^2 \right) = \frac 1 {\sqrt 2} \left(\psi_{2x - 1} \pm \psi_{2x} \right).
}
These operators will be used to define the notion of chiral symmetry later in this section. They satisfy the same anticommutation relations as the spinless fermions $\psi_v$.

To motivate the above definition, consider a very simple theory of spinless fermions given by \cite{Susskind:1976jm}
\bel{\label{def H Susskind}
  H = \sum_{v = 1}^N \i \left(\psi_v\+ \psi_{v + 1} - \psi_{v + 1}\+ \psi_v \right), \quad \psi_{N + 1} \equiv \psi_1.
}
Expressing $H$ in the chiral basis \eqref{def chiral basis} and dropping higher-derivative terms gives the Dirac Hamiltonian, which has a nice (though not needed in what follows) expression in $\gamma$-matrix notation:
\bel{\label{def H Dirac}
  H\_{Dirac}\^{(IR)} = \sum_{x = 1}^{N/2} \i \left((\Psi^+_x)\+ \del \Psi^+_x - (\Psi^-_x)\+ \del \Psi^-_x\right) \equiv \sum_x \i \bar \Psi_x \c \del \Psi_x,
}
with $\del \O_x \equiv \O_x - \O_{x - 1}$. This is the formulation of spinors as \emph{staggered fermions} \cite{Kogut:1974ag, Susskind:1976jm}. The decoupled fermions $\Psi^\pm_x$ are identified with right- and left-moving modes; they satisfy the Dirac equation with opposite velocities,
\bel{
  \dot \Psi^\pm_x \equiv \i\, [H\_{Dirac}\^{(IR)}, \Psi^\pm_x] = \pm \del \Psi^\pm_x.
}

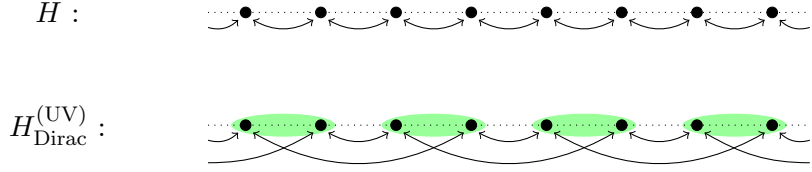
\begin{figure}
\begin{center}

\begin{tikzpicture}[scale = 2]

  \draw[step = 0.5, dotted] (-2, 0) -- (2, 0);

  \foreach \x in {-1.75, -1.25, ..., 1.75} {
    \filldraw[black] (\x, 0) circle (1pt);
  };

  \foreach \x in {-1.75, -1.25, ..., 1.25}
    \draw[<->] (\x + 0.05, -0.05) to[bend right] (\x + 0.45, -0.05);

  \draw[->] (-2, -0.1) to[bend right] (-1.8, -0.05);
  \draw[->] (2, -0.1) to[bend left] (1.8, -0.05);

  \draw (-3, 0) node {$H$\ :};

  \draw (-3, -0.75) node {$H\_{Dirac}\^{(UV)}$\ :};

  \foreach \x in {-1.5, -0.5, 0.5, 1.5} {
    \filldraw[fill = green!40!white, draw = white] (\x, -0.75) ellipse (0.35 and 0.08);
  };

  \draw[step = 0.5, dotted] (-2, -0.75) -- (2, -0.75);

  \foreach \x in {-1.75, -1.25, ..., 1.75} {
    \filldraw[black] (\x, -0.75) circle (1pt);
  };

  \foreach \x in {-1.25, -0.25, 0.75} {
    \draw[<->] (\x + 0.05, -0.8) to[bend right] (\x + 0.45, -0.8);
    \draw[<->] (\x - 0.45, -0.8) to[bend right] (\x + 0.95, -0.8);
  };

  \draw[->] (-2, -0.85) to[bend right] (-1.8, -0.8);
  \draw[->] (2, -0.85) to[bend left] (1.8, -0.8);
  \draw[->] (-2, -1) to[out = 0, in = 210] (-1.3, -0.8);
  \draw[->] (2, -1) to[out = 180, in = -30] (1.3, -0.8);

\end{tikzpicture}
\end{center}
\caption{\small Hoppings between sites in the Hamiltonians \eqref{def H Susskind} and \eqref{def H Dirac prime}. Green ellipses denote sites that combine into a single spinor.}
\label{fig hopping}
\end{figure}

The lattice version \eqref{def H Dirac} of the Dirac Hamiltonian is not Hermitian. The two-derivative terms, which were dropped in anticipation of a long-wavelength approximation, are needed to restore Hermiticity. Indeed, in the chiral basis, the full expression for the starting Hamiltonian \eqref{def H Susskind} is
\bel{
  H = H\_{Dirac}\^{(IR)} - \sum_x \frac \i 2 \left(\del(\Psi_x^+ - \Psi_x^-)\+ \del(\Psi_x^+ + \Psi_x^-) \right).
}
The Hermitian two-derivative term $\frac \i 2 (\del \Psi_x^+)\+ \del \Psi_x^- - \frac \i 2(\del \Psi_x^-)\+ \del \Psi_x^+$ is the only term in $H$ that mixes $\Psi^+$ and $\Psi^-$ modes. An acceptable Hermitian operator that reduces to the Dirac theory at low momenta and that does not mix chiralities at any momentum scale is obtained by dropping this two-derivative term from $H$, or equivalently by taking the Hermitian part of $H\_{Dirac}\^{(IR)}$. This gives
\bel{\label{def H Dirac prime}
  H\_{Dirac}\^{(UV)} \equiv H\_{Dirac}\^{(IR)} - \sum_x \frac \i 2 \left( (\del \Psi_x^+)\+ \del \Psi_x^+ - (\del \Psi_x^-)\+ \del \Psi_x^- \right).
}
This Hamiltonian secretly describes  two decoupled spinless fermions hopping on disjoint sublattices. This can be seen by expressing it in terms of original fermion variables (fig.~\ref{fig hopping}):
\bel{
  H\_{Dirac}\^{(UV)} = \sum_x \frac{\i}2 \left( \psi_{2x}\+ \psi_{2x + 1} - \psi_{2x + 1}\+ \psi_{2x} \right) + \sum_x \frac{\i}2 \left( \psi_{2x-3}\+ \psi_{2x} - \psi_{2x}\+ \psi_{2x - 3} \right).
}

Working with a Hamiltonian like $H$ in \eqref{def H Susskind} is one of the most natural ways to avoid the well-known ``doubling problem,'' i.e.~to obtain a lattice theory of a $d = 1$ spinor with a linear, relativistic dispersion around a \emph{single} point in momentum space. The crucial point here is that $H$ \emph{must} couple the modes of different chiralities. If these modes are decoupled, the theory may not be Hermitian, as is the case of $H\_{Dirac}\^{(IR)}$ in \eqref{def H Dirac}, or each mode must ``double'' and have two zero-energy modes on its own, which is what happens with $H\_{Dirac}\^{(UV)}$ \cite{Nielsen:1980rz, Nielsen:1981xu, Friedan:1982nk}. If modes of different chiralities are decoupled so that their numbers are separately conserved, the theory has \emph{chiral symmetry}.

The \emph{chiral anomaly} is logically distinct from fermion doubling. A theory may have chiral symmetry regardless of how many Dirac points exist in momentum space; topological considerations only require that there be an even number of such points. This chiral symmetry may or may not be anomalous, and establishing a criterion for this anomaly is one goal of this paper. To this end, a more precise definition of chiral symmetry (and some related concepts) must be introduced.


In analogy with the ordinary fermion number $n_v = \psi\+_v \psi_v$, \emph{chiral fermion numbers} are defined  as
\bel{
  n_x^\pm \equiv (\Psi^\pm_x)\+ \Psi^\pm_x.
}
In terms of the bilinears in \eqref{def Z S}, these operators are nicely recorded as
\algns{
  n_x \equiv n_x^+ + n_x^- &=  n_{2x - 1} + n_{2x} = \1 - \frac12 (Z_{2x - 1} + Z_{2x}) ,\\
  n_x\^A \equiv n_x^+ - n_x^- &=  \psi\+_{2x - 1} \psi_{2x} + \psi\+_{2x} \psi_{2x - 1} = \frac12 (S_{2x - 1, 2x} + S_{2x, 2x - 1}).
}
``A'' stands for ``axial,'' a word synonymous with ``chiral'' in this paper.  All pairs of $n_x$ and $n_y\^A$ commute, even when $x = y$. In the $4\times 4$ subspace on the dimer $x$, they are given by
\bel{\label{def n nA}
  n_x = \left[
                  \begin{array}{cccc}
                    0 & 0  & 0 & 0 \\
                    0 & 1  & 0 & 0 \\
                    0 & 0  & 1 & 0 \\
                    0 & 0  & 0 & 2 \\
                  \end{array}
                \right], \quad
  n_x\^A = \left[
                  \begin{array}{cccc}
                    0 & 0  & 0 & 0 \\
                    0 & 0  & 1 & 0 \\
                    0 & 1  & 0 & 0 \\
                    0 & 0  & 0 & 0 \\
                  \end{array}
                \right].
}
Technically, there is an ordering ambiguity in defining the state obeying $n_{2x - 1} = n_{2x} = \1$, i.e.~the state with a fermion on each site of the dimer, but this does not affect the two matrices above. Note that $n_x$ measures the effective fermion number, while $n_x\^A$ measures the \emph{spin} of the spinor state. The spin is vanishing when $x$ has zero or two fermions: exactly one fermion must be on $x$ for it to spin (i.e.~for it to oscillate between the two sites on the dimer), and in the one-fermion subspace the spin $n\^A_x$ has eigenvalues $\pm1$.

The total fermion parity is the operator
\bel{\label{def -1F}
  (-1)^F \equiv \prod_v (-1)^{n_v}.
}
It generates a kinematic $\Z_2$ symmetry of the system. In other words, $(-1)^F$ is a $\Z_2$ operator that generates the center of the algebra of all fermion bilinears. In particular, this means it commutes with any Hamiltonian built out of these bilinears. This is an \emph{on-site} symmetry, which means that its generator is a product of local operators, and so it can be gauged by placing $\Z_2$ gauge fields on all the links. It is also possible to gauge its coarse-grained, \emph{off-site} version, by placing gauge fields on links between dimers and imposing the Gauss law at each dimer instead of at each site. This is an important difference that will be discussed in more detail in the next section.

Note the local relation
\bel{\label{spin-charge}
  (-1)^{n_x} = (-1)^{n_x\^A}.
}
This can be understood as a rudimentary \emph{spin-charge relation} \cite{Wen:2013ue, Seiberg:2016gmd, Seiberg:2016rsg}: spinor states with odd fermion number have a spin of nonzero magnitude, and spinors with even fermion number have zero spin. Because of this, $(-1)^F$ also measures the total spin parity in the system, which is thus kinematically conserved.

Finally, it is useful to recall that gauging the $\Z_2$ fermion parity in a spinless fermion theory is done by placing a $\Z_2$ gauge field on each link and imposing $X_{v - \frac12} Z_v X_{v + \frac12} = \1$. Operators in the $\Z_2$ gauge theory are Pauli matrices on links, and will be denoted $X_{v + \frac12}$, $Z_{v + \frac12}$, and so on; a link between $v$ and $v + 1$ will be denoted $v + \frac12$. The algebra of gauge-invariant operators is generated by
\bel{
  X_{v + \frac12}, \quad Z_v, \quad \~S_{v, v+1} \equiv Z_{v + \frac12} S_{v, v + 1}.
}
While $n_x$ remains gauge-invariant, $n_x\^A$ must be replaced by
\bel{\label{def tilde nA}
  \~n_x\^A \equiv Z_{2x - \frac12} n_x\^A = \frac12 \left(\~S_{2x - 1, 2x} + \~S_{2x, 2x - 1} \right).
}
Note that the spin-charge relation \eqref{spin-charge} gets modified to
\bel{\label{spin-charge mod}
  (-1)^{n_x} = (-1)^{\~n\^A_x}.
}
This relation will make it possible to couple the electric field to the spin of a Dirac spinor. 

The fact that $n\^A_x$ is not gauge-invariant lies at the heart of the chiral anomaly in this finite context. However, before elaborating on that, it is necessary to examine in more detail some additional symmetries of the Dirac theory.

\section{The curious incident of the $\mathrm U(1)$ symmetry in the Dirac theory} \label{subsec curious}

The Dirac theory \eqref{def H Dirac} and its UV completion \eqref{def H Dirac prime} have two important quantities conserved due to \emph{off-site} symmetries: the total fermion number and the total spin (i.e.~the \emph{axial number}). They are respectively given by
\bel{\label{def NF NA}
  N\^F \equiv \sum_x n_x = \sum_v n_v, \quad
  N\^A \equiv \sum_x n_x\^A.
}
The claim that the symmetries associated with these conserved quantities are not on-site is not standard, and may appear surprising. The purpose of this section is to justify it.

To start off, consider the fermion number $N\^F$. The following short argument offers a preview of the issues at hand. Standard lore has it that U(1) is the group associated to fermion number symmetry, since an $N\^F$-conserving Hamiltonian such as \eqref{def H Susskind} is invariant under $\psi_v \mapsto \e^{\i\theta} \psi_v$. What is the generator of this symmetry? A natural guess is $\e^{\i N\^F \d\theta}$, for an infinitesimal $\d\theta$; indeed, it is easy to see that $\e^{-\i N\^F \d\theta} \psi_v\, \e^{\i N\^F \d\theta} = \e^{\i\d\theta} \psi_v$. As such, this symmetry appears to be on-site, its generator being equal to $\prod_v \e^{\i n_v \d\theta}$. Given this on-site presentation, the natural next step is to try to gauge this symmetry by placing U(1) gauge fields $A_{v + \frac12}$ on all links and imposing the gauge constraint
\bel{\label{aux gauge constraint}
  L_{v - \frac12} \e^{\i n_v \d\theta} L\+_{v + \frac12} = \1,
}
where electric field operators act as $L_{v + \frac12}\qvec{A_{v + \frac12}} = \qvec{A_{v + \frac12} + \d\theta}$.%
\footnote{Expanding in $\d \theta$ gives a more conventional form of the Gauss law, $n_v = \del E_{v + \frac12}$, with the usual definition of the U(1) electric field $E = \i \pder{}A $. The form given in \eqref{aux gauge constraint} is more general: it also applies to finite gauge groups.

On a related note, in field theory, the global charge (like $N\^F$ here) is often also called the generator of a symmetry. In this paper, the term ``symmetry generators'' will be reserved for operators that are also generating elements of the symmetry group. With this convention, symmetry generators are necessarily invertible, unlike $N\^F$, and the symmetry group they generate spans (a subalgebra of) the double commutant of the Hamiltonian.} %
Consider the product of the Gauss law \eqref{aux gauge constraint} over all sites $v$. The resulting ``singlet constraint'' that matter fields must obey is $\e^{\i N\^F \d\theta} = \1$. Since $\d \theta$ is arbitrarily small, and since eigenvalues of $N\^F$ lie between $0$ and $N$ (the fixed number of sites on the lattice), the singlet constraint imposes $N\^F = 0$. Thus gauging the U(1) eliminates all matter excitations, which is not \emph{quite} what a reasonable gauge theory should do.%
\footnote{This argument has nothing to do with the dimensionality of the system or the fact that gauge theories are essentially trivial in $d = 1$, as the focus is on the singlet relations in the matter sector alone.} %
A symmetry whose gauging restricts the matter to just one state will be called \emph{quasi-anomalous}.

Quasi-anomalous symmetries should be contrasted to anomalous ones. No state can simultaneously satisfy all local constraints when an anomalous symmetry is gauged. When quasi-anomalous symmetries are gauged, on the other hand, there are still gauge degrees of freedom constrained in the usual way: all closed electric flux lines atop the $N\^F = 0$ matter state are allowed. This is thus a special situation in which gauge fields decouple from the matter. This effect is essentially \emph{classical}: the quasi-anomaly, just like the actual anomaly to be discussed below, does not arise from the noncommutativity of Gauss operators \eqref{aux gauge constraint}, which would have made it intrinsically quantum.


It is instructive to further explore the onset of quasi-anomalous behavior. An intuitive attempt to stop gauging from killing off all matter degrees of freedom would be to consider different generators of the particle number symmetry. For instance, consider $\prod_v \e^{\i (n_v - m) \d\theta}$ for some $m \in \R$. Conjugating $\psi_v$ by this U(1) generator again results in $\e^{\i \, \d\theta} \psi_v$. The corresponding Gauss law is
\bel{\label{aux gauge constraint 2}
  L_{v - \frac12} \e^{\i (n_v - m) \d\theta} L\+_{v + \frac12} = \1,
}
and it enforces the singlet constraint $N\^F = m N$. If $m = 1/2$, gauging will restrict the matter to states with exactly $N/2$ fermions, so this appears to be a rather natural way to impose the constraint that physical states are exactly at half-filling. However, U(1) is a compact group, and so there must exist an (infinitely large) number $K \equiv \frac{2\pi}{\d\theta}$ that satisfies $L_{v + \frac12}^K = \1$. Raising either the Gauss law \eqref{aux gauge constraint 2} or  the generator of U(1) to the $K$'th power gives $\e^{- 2\pi \i m} = \1$, which can only be fulfilled for integer $m$. Thus $m = 0, 1$ are the only values for which there exist \emph{any} states obeying the singlet constraint. In either case, all matter excitations are again projected out.

Even if $m = 1/2$ had been allowed in \eqref{aux gauge constraint 2}, the resulting singlet constraint would have still been pathological. One way to see this is to consider the most general U(1) symmetry generator,
\bel{\label{def general U1 gen}
  \prod_v \e^{\i q(n_v) \d \theta},
}
for some function $q: \{0, 1\} \rar \Z$. This assigns charge $q(0)$ to a site with no fermion, and charge $q(1)$ to a site with a fermion. If $q(0) = q(1)$, the gauge fields decouple from the matter, so it is enough to focus on $q(0) \neq q(1)$. 
The singlet constraint is
\bel{\label{dumb constraint}
  N\^F = \frac{q(0)}{q(0) - q(1)} N.
}
By choosing, say, $q(n) = (-1)^n$, the ``half-filling'' constraint $N\^F = N/2$ appears again. Other choices of $q(n)$ can give rise to singlet constraints with arbitrary rational ratios $N\^F/N \equiv \xi$.%
\footnote{\label{foot qs} Conjugating by the operator \eqref{def general U1 gen} induces the map $\psi_v \mapsto \e^{\i(q(1) - q(0))\d\theta}\psi_v$. To get $\psi_v \mapsto \e^{\i\d\theta} \psi_v$, it is necessary to set $q(1) - q(0) = 1$, i.e.~$\xi \in \Z$. Assuming this, the constraint \eqref{dumb constraint} can only be fulfilled for $q(0) = 0$ or $q(0) = -1$, i.e.~$\xi = 0$ or $\xi = 1$. The main text will not assume $\xi$ is an integer, as it aims to illustrate a different point.} %

An on-site U(1) symmetry with $\xi = 0$ or $\xi = 1$ is thus quasi-anomalous, and if $\xi < 0$ or $\xi > 1$, the symmetry is anomalous. The situation at $\xi \in (0, 1)$ is more subtle. For fixed $\xi$, the U(1) cannot be gauged for every $N$. E.g.~at ``half-filling,'' for $\xi = 1/2$, any odd $N$ yields a constraint that \emph{no} pure state can satisfy. This should be regarded as an anomaly, albeit one that depends on the underlying geometry (the lattice ``volume'' $N$). For any $N$, some $q(n)$ (and hence $\xi$) can be chosen to define a U(1) that is anomalous because $\xi N$ in the constraint \eqref{dumb constraint} is not an integer. Thus, whenever the particle number U(1) acts on-site on a finite lattice, there exists a theory in which the U(1) is either geometrically anomalous or quasi-anomalous.\footnote{Note, once again, that \emph{no} anomalies mentioned in this passage are quantum anomalies.}


Similar arguments show that $\Z_K$ on-site symmetries at $K > 2$ can always be made quasi-anomalous. Consider the symmetry generated by $\prod_v \e^{\i \frac{2\pi}K q(n_v)}$ for some $q:\{0,1\} \rar \{0, \ldots, K - 1\}$.
Gauging this symmetry results in the ``finite $K$'' version of the singlet constraint \eqref{dumb constraint}
\bel{\label{other dumb constraint}
  \big(q(0) - q(1)\big)\, N\^F = q(0) N \ \trm{mod}\ K.
}
Taking $q(1) = 1$ and $q(0) = 0$ (cf.~footnote \ref{foot qs}), the constraint \eqref{other dumb constraint} can be satisfied by only one state for any $2\leq  N \leq K - 1$, making the $\Z_K$ quasi-anomalous. The U(1) case is obtained by taking $K \gg q(0), N$.

The $\Z_2$ global symmetry is special. There are no impediments to gauging it. Possible singlet constraints are $N\^F = 0 \, \trm{mod}\, 2$, which comes from gauging fermion parity in the usual way, and $N - N\^F = 0\,\trm{mod}\, 2$, which comes from gauging the parity of ``holes'' (sites without a fermion).

These examples show that whenever one tries to establish a $\Z_K$ ($K > 2$) or U(1) on-site fermion number symmetry, it will be possible to pick charges $q(n)$ such that the symmetry is anomalous or quasi-anomalous for some $N$. One way to avoid such situations for all $N$ and $q(n)$ is to view the particle number symmetry as an off-site one. In this paper, this will be done by defining symmetry groups associated to $N\^F$ to be off-site in such a way that \emph{no choice} of charges in the Gauss law can lead to a (quasi-)anomaly, with some mild controlled exceptions:
\begin{center}
\begin{minipage}{0.85\textwidth} \label{rule}
  \emph{Off-site rule}: an Abelian fermion number symmetry must act on blocks that have at least $D$ different possible fermion numbers, where $D$ is the order of the group.
\end{minipage}
\end{center}

A consequence of the off-site rule is that the U(1) symmetry generated by $\e^{\i N\^F \d \theta}$ can \emph{never} be sufficiently off-site to be nonanomalous, unless $N$ is taken to infinity such that it scales with the size of U(1). To get a feeling for what this means in a simpler context, consider some minimally off-site symmetries. One example is the $\Z_2$ generated by
\bel{
  Q_{(2)} \equiv \prod_x (-1)^{n_x}.
}
This generator obeys the off-site rule as there are three possible fermion numbers on each dimer. It is actually equal to the fermion parity $(-1)^F$ from \eqref{def -1F}. If this symmetry were gauged, the corresponding gauge fields would live on links connecting dimers $x$ and $x \pm 1$.  Gauging this ``on-dimer'' $\Z_2$ imposes the constraint $N\^F = 0\, \trm{mod}\, 2$. This has the same global effect as gauging the on-site fermion parity: it projects the matter sector to states with an even number of fermions only. Indeed, this is due to the fact that $(-1)^{n_x}$ acts on-site even within the dimer.%
\footnote{In $d = 1$ gauge theories, global degrees of freedom are all there is to discuss. Thus for \emph{most} intents and purposes, gauging the on-site $\Z_2$ is equivalent to gauging the ``on-dimer'' $\Z_2$. The main difference between these two scenarios is precisely the chiral anomaly, which will be discussed in section \ref{subsec QEDs}. In $d > 1$, making the $\Z_2$ off-site would change the number of local degrees of freedom in the gauged theory by ``coarse-graining'' the plaquettes. In this case, even before discussing anomalies, it is important to specify which $\Z_2$ is gauged.} %

Another example is
\bel{\label{def Q3}
  Q_{(3)} \equiv \prod_x \e^{\i \frac{2\pi}3 (n_x - 1)}.
}
Gauging produces a theory in which $\Z_3$ gauge fields live on links between dimers, and dimers with exactly two fermions are connected by electric flux lines to dimers with exactly zero fermions. This treats particles above and below ``half-filling'' as oppositely charged excitations. The resulting global singlet constraint is $N\^F = \frac N2\,\trm{mod}\,3$, and can be fulfilled whenever $Q_{(3)}$ exists (if $N$ is odd, $Q_{(3)}$ does not exist). The quasi-anomaly for this symmetry appears only at $N = 2$ when the other generator choices, $\prod_x \e^{\i \frac{2\pi}3 (n_x + 1)}$ and $\prod_x \e^{\i \frac{2\pi}3 n_x}$, are used. This is the kind of ``mild'' anomaly that is allowed by the off-site rule: the example with $\Z_{N +1}$ below will make this clearer.

The other interesting on-dimer $\Z_3$ symmetry is generated by
\bel{\label{def Q3 A}
  Q\^A_{(3)} \equiv \prod_x \e^{\i\frac{2\pi}3 n\^A_x}.
}
The chiral symmetry is back in the story! This incarnation will be denoted $\Z_3\^A$. Each factor in this operator measures the total spin on one dimer. Note that $\Z_3$ and $\Z_3\^A$ generators commute.


The \emph{largest} fermion and axial number symmetry groups that obey the off-site rule on $N$ sites are $\Z_{N + 1}$, as $N + 1$ is the number of distinct eigenvalues of $N\^F$ and $N\^A$. The associated generators are
\bel{\label{def Q QA}
  Q \equiv \exp\left\{\frac{2\pi \i}{N + 1} \left(N\^F - \frac N 2\right)\right\}, \quad Q\^A \equiv \exp\left\{\frac{2\pi \i}{N + 1} N\^A\right\}.
}
As written, these generators make sense for even $N$. They view the whole lattice as one ``site.'' After gauging, the gauge field would live on \emph{one} link.

Note that the $\Z_{N + 1}$ in \eqref{def Q QA} is borderline quasi-anomalous, like the $\Z_3$ was for $N = 2$ in \eqref{def Q3}.  The singlet constraint is $N\^F = N/2$, but by changing $q(0)$ it is possible to obtain constraints $N\^F = 0$ or $N\^F = N$, making the symmetry quasi-anomalous. The $\Z_{N + 1}$ is thus at the very crossover between nonanomalous and anomalous versions of fermion number symmetry groups. The strength of the proposed off-site rule for these symmetries lies in the fact that these extremal cases will be the \emph{only} possible quasi-anomalies.

In the continuum limit ($N \gg 1$), the symmetries will approach the usual U(1) and U(1)$\^A$. If $N$ is large, approximately on-site U(1) symmetries can be obtained by splitting the $N$ sites into $N/M$ blocks of $M \gg 1$ sites, and then placing $\Z_{M + 1} \approx \trm U(1)$ gauge fields on links between blocks. More generally, if block sizes are kept finite, ``on-block'' $\Z_{M + 1}$ and $\Z\^A_{M + 1}$ symmetries can be defined in analogy with the $\Z_3$ and $\Z_3\^A$ defined for dimers ($M = 2$ blocks) in eqs.~\eqref{def Q3} and \eqref{def Q3 A}. Their generators, for even $M$, are conveniently defined as
\bel{\label{def QM QMA}
  Q_{(M + 1)} \equiv \exp\left\{\frac{2\pi \i}{M + 1} \left(N\^F - \frac M 2 \right)\right\}, \quad Q\^A_{(M + 1)} \equiv \exp\left\{\frac{2\pi \i}{M + 1} N\^A\right\}.
}
\newpage

\section{Lattice QEDs and their chiral anomalies} \label{subsec QEDs}

The definition of staggered fermions solved the doubling problem by weaving Dirac spinors into the spatial lattice, effectively assigning different momenta to different components of the spinor field. It is clear that one now has to think carefully about coupling lattice spinors to gauge fields. For instance, do different components of a spinor feel a different gauge field? The approach here will be practical: if a system has a symmetry, then one can talk about gauging it, and gauge fields will live between those sites (or regions) on which the symmetry group acts without (quasi-)anomaly, as discussed in the previous section. It is necessary to understand whether a symmetry of the spinor theory can be thought of as on-site --- and if not, how off-site is it? The off-site rule from section \ref{subsec curious} answer these questions in the following way.

The \emph{chiral symmetry} is never defined on-site. The minimally off-site version of chiral symmetry is the ``on-dimer'' $\Z_3\^A$ generated by $Q_{(3)}\^A$ in \eqref{def Q3 A}.\footnote{There is also the ``on-dimer'' $\Z_2\^A$ generated by $\prod_x (-1)^{n_x\^A}$, but by the spin-charge relation \eqref{spin-charge} this $\Z_2\^A$ is identical to the fermion parity $\Z_2$ generated by $(-1)^F$.} Further, there is a family of less and less on-site chiral symmetries generated by the charges $Q\^A_{(M + 1)}$ in \eqref{def QM QMA} for all even divisors $M$ of $N$ between $2$ and $N$, culminating at $M = N$ with the maximal axial symmetry $\Z_{N + 1}\^A$ defined in \eqref{def Q QA}.

The on-site version of the \emph{fermion number symmetry} is simply the fermion parity, generated by $(-1)^F$. More off-site versions of this symmetry are the $\Z_{M + 1}$'s defined for all divisors $M$ of $N$ in the previous section, again culminating in the maximal particle number symmetry $\Z_{N + 1}$ that is generated by $Q$ in \eqref{def Q QA}.

Lattice quantum electrodynamics (QED) is, in principle, any theory obtained by coupling complex spinors to Abelian gauge fields, i.e.~by gauging a symmetry related to fermion number. For a given fermion system, e.g.~\eqref{def H Susskind} or \eqref{def H Dirac prime}, there exist many QEDs that differ in how off-site the gauged symmetry is. The most rudimentary of these is the $\Z_2$ QED, in which fermion parity is gauged, there is a $\Z_2$ gauge field on each link, and different spinor components indeed feel different gauge fields. The operators in this theory were defined at the end of section \ref{subsec defs}, where it was remarked that $n_x\^A$ was not gauge-invariant, but that its substitute $\~n_x\^A$ in \eqref{def tilde nA} was. In particular, this means that gauging the $(-1)^F$ parity now couples the two chiral modes and allows the chiral symmetry to be broken. For instance, the maximal gauge-invariant axial symmetry generator is
\bel{
  \~Q\^A \equiv \exp\left\{\frac{2\pi \i}{N + 1} \sum_x \~n_x\^A \right\},
}
and it does not commute with $X_{2x - \frac12}$, the electric field between the two spinor components. An electric field operator acts on the Wilson line connecting two fermion operators within $\~n_x\^A$, and thereby effects a change of the total spin of  the dimer (see the discussion of the modified spin-charge relation \eqref{spin-charge mod}).

As a sharper diagnostic of this breakdown of axial number conservation, consider turning on an electric field, i.e.~evolving with the Hamiltonian $g^2 \sum_v X_{v + \frac12}$. The change in the axial number is
\bel{\label{dN dt}
  \der{\~N\^A}t = \i g^2 \sum_{v = 1}^N \left[ X_{v + \frac12}, \~N\^A\right] =  2 \i g^2 \sum_{x = 1}^{N/2}  X_{2x - \frac12} \~n\^A_x.
}
Applying an electric field to a translation-invariant $\~N\^A$-eigenstate thus induces a change of the axial number proportional to $2g^2 \sum_x X_{2x - \frac12} \d t$. The analogy with the familiar field theory effect --- the axial current nonconservation proportional to $\int \d^2 x\,  \epsilon^{\mu\nu} F_{\mu\nu} = 2\int \d^2 x \, E_1(x)$ --- makes it natural to refer to this $\~N\^A$ nonconservation as the chiral anomaly. This is one of the main results of this paper. The anomalous chiral symmetry group here can range from $\Z_3\^A$ in \eqref{def Q3 A} to $\Z_{N + 1}\^A$ in \eqref{def Q QA}.

There are more exotic setups in which the chiral anomaly can manifest itself. Consider gauging the $\Z_{M + 1}$ version of particle number symmetry, with $M$ being an \emph{odd} divisor of $N$. Then some of the dimers used to define spinors must also carry a $\Z_{M + 1}$ gauge field. The corresponding axial numbers $n_x\^A$ are thus not gauge-invariant, and their gauge-invariant versions $\~n_x\^A$ will again fail to commute with electric fields, ruining the conservation of $\~N\^A$.

Conversely, some off-site gaugings of particle number do not lead to chiral anomalies. In particular, the on-dimer $\Z_2$ or $\Z_3$ can be gauged with impunity. All gauge fields will live between dimers, and all $n_x\^A$ will be gauge-invariant. Thus, there exist lattice formulations of QED both with and without a chiral anomaly. The \emph{main criterion} is whether a link connecting two components of the same spinor also contains a gauge field associated to particle number gauging.

The first example of the chiral anomaly given here --- the $\Z_2$ QED --- is a particularly natural situation. This $\Z_2$ gauge field does not merely gauge the fermion parity: it also acts as a dynamical \emph{spin structure} \cite{Radicevic:2018okd}. It is a necessary ingredient in any bosonization of the fermion theory \cite{Gaiotto:2015zta, Chen:2017fvr, Chen:2018nog} (see also \cite{Thorngren:2018bhj} for another perspective on anomalies in this context). This is important because bosonization, by summing over spin structures, precisely removes all the ordering ambiguities that naturally appear in the definition of a fermion many-body Hilbert space (a cursory mention of this issue was made below eq.\ \eqref{def n nA}). If a link did not contain a $\Z_2$ gauge field, the fermion creation operators on the edges of this link would have an ordering ambiguity. Therefore all links must have a $\Z_2$ field in order to remove the ambiguities.

One might want to gauge a bigger symmetry by placing $\Z_{M + 1}$ gauge fields on certain links only. The existence of a bosonic dual still requires $\Z_2$ gauge fields on all links, however. If the fermion parity $\Z_2$ is a subgroup of $\Z_{M + 1}$, i.e.~if the $\Z_{M + 1}$ gauging also enforces $N\^F = 0 \,\trm{mod}\, 2$, then the ambiguity-free theory should more properly be recorded as a $\Z_2 \ltimes (\Z_{M + 1}/\Z_2)$ gauge theory, with the first factor being a gauged on-site symmetry, and the second factor being a gauged off-site symmetry. (If $M \gg 1$, the gauge group should be understood as $\Z_2 \ltimes\trm U(1)$, with only $\Z_2$ acting on-site.) If $\Z_2$ is not a subgroup of $\Z_{M + 1}$, the gauge group of the ambiguity-free theory is $\Z_2 \times \Z_{M + 1}$.

\section{Remarks} \label{sec remarks}

A number of elementary but nonstandard claims were made in this paper. There are two main lessons: finite systems can exhibit a chiral anomaly, and they cannot have continuous symmetry groups. The second point, more precisely, claims that if the off-site rule is violated, novel geometric anomalies and quasi-anomalies of on-site symmetries will appear.

All of these results are purely kinematical and do not depend on the Hamiltonian, as long as particle number is conserved. Further, even though the phrase ``chiral anomaly'' is most pertinent in situations when gauging the particle number symmetry ruins the conservation of the axial number that held in the ungauged theory (e.g.\ in \eqref{def H Dirac prime}), the effect in eq.~\eqref{dN dt}, namely the change of $\~N\^A$ under an electric field, exists even in theories with no chiral symmetry in the UV (e.g.\ in \eqref{def H Susskind}).

The anomalies in section \ref{subsec curious} \emph{do not} necessarily invalidate the vast body of work done on lattice systems. Gauging a quasi-anomalous symmetry is essentially identical to \emph{quenching} fermions in lattice QED \cite{Rothe:1992nt}, which is already an often used (though uncontrolled) approximation in  numerical simulations. In more analytical approaches, a coarse-graining may have been implicitly used whenever a U(1) particle number symmetry was gauged in a lattice theory. While this was mentioned around \eqref{def QM QMA}, it may be useful to illustrate it very explicitly. For a divisor $M$ of $N$, define 
\bel{\label{def tilde psi}
  \~\psi_b \equiv \frac 1 {\sqrt M}  \sum_{i = 1}^{M} \psi_{(b - 1)M + i}, \quad b = 1, \ldots, \frac NM.
}
These coarsened fermions behave much like the original ones, e.g.\ they satisfy $\left\{\~\psi_b, \~\psi_{b'}\+\right\} = \delta_{bb'}$ and $\~\psi_b^2 = 0$. However, they support higher occupation numbers per block $b$, with the number operator
\bel{
  n_b \equiv \~\psi_b\+ \~\psi_b = \frac1M \sum_{i = 1}^M \psi\+_{(b - 1)M + i} \psi_{(b - 1)M + i}
}
having $M + 1$ distinct eigenvalues. Thus one can now write a Hamiltonian like
\bel{
  \~H = \sum_b \left(\~\psi_b\+ \~\psi_{b + 1} + \~\psi_{b + 1}\+ \~\psi_b + V(n_b)\right),
}
and claim that it has a $\Z_{M + 1}$ ''on-block'' symmetry. In fact, $M \rar \infty$ can be taken without affecting the appearance of $\~H$, but the dimension of the Hilbert space diverges as $\sim\e^M$ in this limit. No fermion system with such a U(1) symmetry can have a finite-dimensional Hilbert space before coarse-graining. It would be interesting to revisit the discussions of on-site symmetries and anomalies (starting from \cite{Chen:2011pg, Wen:2013oza, Kapustin:2014tfa}) in the context of this new understanding. The geometric \hbox{(quasi-)anomalies} of U(1) appear to be ``curable'' by anomaly inflow \cite{Callan:1984sa}, just like their quantum counterparts. This discussion, similar in spirit to the one in \cite{Komargodski:2017smk}, will appear elsewhere.

Geometric anomalies of on-site symmetries may also be detected in the ``classical'' continuum limit, i.e.~in a regime where the lattice spacing is taken to be infinitesimal and dimensionful versions of quantum fields are used, without necessarily tuning to a critical point. Consider a theory of spinless fermions on $N$ sites, with $\Z_K$ gauge fields on all links, and a gauge constraint of the general form
\bel{
  L_{v - \frac12} \e^{\i \frac{2\pi}K q(n_v)} L\+_{v + \frac12} = \1.
}
A continuum theory of U(1) gauge fields is obtained by sending $N, K \rar \infty$. The standard way to control this limit is via an infinitesimal lattice spacing $a$, which is used to define the continuum volume $V$ and the continuum fields $A\_{cont}(v)$ and $\psi\_{cont}(v)$ as
\bel{
  N \equiv \frac V a, \quad \exp\left\{\i \frac{2\pi}K n_{v + \frac12}\right\} \equiv \e^{\i a A\_{cont}(v)}, \quad \psi\_{cont}(v) \equiv \frac1{\sqrt a} \psi_v.
}
In particular, a standard assumption when working with continuum gauge fields is that $A\_{cont}(v)$ has a continuous spectrum, meaning that $Ka \rar \infty$. Thus, if the continuum volume $V$ is finite, $K \gg N \gg 1$ must be assumed in order to get the familiar continuum limit. This is also the regime in which the geometric anomalies can be detected, as the off-site rule is violated.

Indeed, the singlet constraint in this limit can be written as
\bel{
  0 = \sum_v q(n_v) = \int_0^V \d v \, \rho(v),
}
where $\rho(v) \equiv \frac 1a q(n_v)$ is the continuum version of the charge density.\footnote{As defined here, $\rho(v)$ is not an integrable function. The integrals here must be understood in a formal sense. They are given more conventional meanings only after coarse-graining, i.e.~if the continuum limit is reached by letting $N/M \rar \infty$ in the theory of effective fermions $\~\psi_b$ in \eqref{def tilde psi}. Rescaling by $\sqrt{\~ a} = \sqrt{Ma}$ gives continuum coarse-grained fermion fields $\~\psi\_{cont}(b)$, which is what is usually meant by a continuum fermion $\psi(x)$.} Now consider, for example, $q(n_v) = 2n_v - 1$, which is the charge assignment that leads to the half-filling constraint $N\^F = \frac12 N$. Here the continuum charge density $\rho(v)$ has an infinite negative contribution diverging as $1/a$, and the singlet constraint in terms of the continuum fields is
\bel{
  \int_0^V \d v \ \psi\_{cont}\+(v) \psi\_{cont}(v) = \frac12 \frac V a.
}
A consistent singlet constraint is possible only if $V/a$ is an even integer. This kind of detail is typically lost when the continuum limit is taken, with multiplicative constants in front of $V/a$ being considered nonuniversal from the point of view of the continuum theory. However, the lesson from this example is that additional care must be taken when regularizing the theory: some regularizations are inconsistent due to geometric anomalies. Far from being unimportant, nonuniversal terms of the form $V/2a$ contain information that diagnoses whether the gauge theory with given matter charges can exist at all.

The analysis presented here applies in higher dimensions, too. The chiral fermion numbers do not have a natural definition for even $d$, but in any dimension the  argument of section \ref{subsec curious} shows that fermion systems can only have discrete on-site symmetries free of (quasi-)anomalies. Chiral symmetry (and a corresponding anomalous nonconservation of the axial number) can be demonstrated in all odd $d$, despite the fact that staggered fermion constructions of Dirac spinors do not fully solve the fermion doubling problem in $d > 1$ \cite{Susskind:1976jm}.

Taking a broader view, the $\Z_2$ QED anomaly presented here may be understood as the analogue of both the chiral and the gravitational anomaly in $d = 1$ systems. The anomaly due to gauging fermion parity was presented as an Adler-Bell-Jackiw anomaly \cite{Adler:1969, Bell:1969} in \eqref{dN dt}, with the $\Z_2$ gauge fields playing lattice analogues of continuum U(1) fields. However, once the $\Z_2$ gauge field is interpreted as a spin structure, it can also be viewed as the lattice version of the $\Z_2$ component of the spin connection. Then one can consider a chiral fermion theory with fixed $N\^A \neq 0$; in this theory, the existence of the gravitational anomaly \cite{AlvarezGaume:1983ig} is reflected by the fact that the $\Z_2$ gauge fields cannot be turned on (i.e.~the spin structure cannot be made dynamical) while keeping constant the nonzero expectation value of $N\^A$, which is not a gauge-invariant operator. This dual role of $\Z_2$ gauge fields can be interpreted as another formulation of the lattice spin-charge relations \eqref{spin-charge} and \eqref{spin-charge mod}.

A point that deserves further elaboration is the fate of nonabelian color or flavor symmetry in manifestly finite systems. For instance, the arguments given here show that a system of two complex fermions cannot have an U(2) on-site flavor symmetry, while no such restrictions appear for an SU(2) symmetry. Presumably the U(1) component of U(2) must emerge upon coarse-graining, like the particle number U(1) did. Understanding this, and generalizing to arbitrary nonabelian symmetries, remains a question for future work.

\section*{Acknowledgments}

I would like to thank Davide Gaiotto, Tarun Grover, Wen  Wei Ho, Chao-Ming Jian, Anton Kapustin, Zohar Komargodski, Steve Shenker, and Xiao-Gang Wen for useful conversations. I thank Weicheng Ye for carefully proofreading the manuscript. This research was supported in part by the National Science Foundation under Grant No.\ NSF PHY-1748958, and I thank the Kavli Institute for Theoretical Physics at UC Santa Barbara for providing the stimulating environment during which most of this work was done. Research at Perimeter Institute is supported by the Government of Canada through Industry Canada and by the Province of Ontario through the Ministry of Economic Development \& Innovation.

\bibliographystyle{ssg}
\bibliography{AnomaliesRefs}

\begingroup\raggedright\begin{thebibliography}{10}

\bibitem{Weinberg:1996kr}
S.~Weinberg, {\em {The quantum theory of fields. Vol. 2: Modern applications}}.
\newblock Cambridge University Press, 2013.

\bibitem{Harvey:2005it}
J.~A. Harvey, ``{TASI 2003 lectures on anomalies},'' 2005.
\newblock \href{https://arxiv.org/abs/hep-th/0509097}{{\tt hep-th/0509097}}.

\bibitem{Adler:1969}
S.~L. {Adler}, ``{Axial-Vector Vertex in Spinor Electrodynamics},'' {\em
  Physical Review} {\bf 177} (1969) 2426--2438.

\bibitem{Bell:1969}
J.~S. {Bell} and R.~{Jackiw}, ``{A PCAC puzzle: {$\pi^0 \rar \gamma \gamma$} in
  the {$\sigma$}-model},'' {\em Nuovo Cimento A Serie} {\bf 60} (1969) 47--61.

\bibitem{Fujikawa:1979ay}
K.~Fujikawa, ``{Path Integral Measure for Gauge Invariant Fermion Theories},''
  {\em Phys. Rev. Lett.} {\bf 42} (1979) 1195--1198.

\bibitem{Ambjorn:1983hp}
J.~Ambj{\o}rn, J.~Greensite, and C.~Peterson, ``{The Axial Anomaly and the
  Lattice Dirac Sea},'' {\em Nucl. Phys.} {\bf B221} (1983) 381--408.

\bibitem{Cho:2014jfa}
G.~Y. Cho, J.~C.~Y. Teo, and S.~Ryu, ``{Conflicting Symmetries in Topologically
  Ordered Surface States of Three-dimensional Bosonic Symmetry Protected
  Topological Phases},'' {\em Phys. Rev.} {\bf B89} (2014), no.~23 235103,
  \href{https://arxiv.org/abs/1403.2018}{{\tt 1403.2018}}.

\bibitem{Kapustin:2014zva}
A.~Kapustin and R.~Thorngren, ``{Anomalies of discrete symmetries in various
  dimensions and group cohomology},''
  \href{https://arxiv.org/abs/1404.3230}{{\tt 1404.3230}}.

\bibitem{Kogut:1974ag}
J.~B. Kogut and L.~Susskind, ``{Hamiltonian Formulation of Wilson's Lattice
  Gauge Theories},'' {\em Phys. Rev.} {\bf D11} (1975) 395--408.

\bibitem{Susskind:1976jm}
L.~Susskind, ``{Lattice Fermions},'' {\em Phys. Rev.} {\bf D16} (1977)
  3031--3039.

\bibitem{Wen:2013oza}
X.-G. Wen, ``{Classifying gauge anomalies through symmetry-protected trivial
  orders and classifying gravitational anomalies through topological orders},''
  {\em Phys. Rev.} {\bf D88} (2013), no.~4 045013,
  \href{https://arxiv.org/abs/1303.1803}{{\tt 1303.1803}}.

\bibitem{Wen:2013ue}
X.-G. Wen, ``{Symmetry-protected topological invariants of symmetry-protected
  topological phases of interacting bosons and fermions},'' {\em Phys. Rev.}
  {\bf B89} (2014), no.~3 035147, \href{https://arxiv.org/abs/1301.7675}{{\tt
  1301.7675}}.

\bibitem{Seiberg:2016gmd}
N.~Seiberg, T.~Senthil, C.~Wang, and E.~Witten, ``{A Duality Web in 2+1
  Dimensions and Condensed Matter Physics},'' {\em Annals Phys.} {\bf 374}
  (2016) 395--433, \href{https://arxiv.org/abs/1606.01989}{{\tt 1606.01989}}.

\bibitem{Seiberg:2016rsg}
N.~Seiberg and E.~Witten, ``{Gapped Boundary Phases of Topological Insulators
  via Weak Coupling},'' {\em PTEP} {\bf 2016} (2016), no.~12 12C101,
  \href{https://arxiv.org/abs/1602.04251}{{\tt 1602.04251}}.

\bibitem{Radicevic:2018okd}
{\DJ}.~Radi{\v c}evi{\'c}, ``{Spin Structures and Exact Dualities in Low
  Dimensions},'' \href{https://arxiv.org/abs/1809.07757}{{\tt 1809.07757}}.

\bibitem{AlvarezGaume:1983ig}
L.~Alvarez-Gaume and E.~Witten, ``{Gravitational Anomalies},'' {\em Nucl.
  Phys.} {\bf B234} (1984) 269.

\bibitem{Bardeen:1984pm}
W.~A. Bardeen and B.~Zumino, ``{Consistent and Covariant Anomalies in Gauge and
  Gravitational Theories},'' {\em Nucl. Phys.} {\bf B244} (1984) 421--453.

\bibitem{Nielsen:1980rz}
H.~B. Nielsen and M.~Ninomiya, ``{Absence of Neutrinos on a Lattice. 1. Proof
  by Homotopy Theory},'' {\em Nucl. Phys.} {\bf B185} (1981) 20.

\bibitem{Nielsen:1981xu}
H.~B. Nielsen and M.~Ninomiya, ``{Absence of Neutrinos on a Lattice. 2.
  Intuitive Topological Proof},'' {\em Nucl. Phys.} {\bf B193} (1981) 173--194.

\bibitem{Friedan:1982nk}
D.~Friedan, ``{A Proof of the Nielsen-Ninomiya Theorem},'' {\em Commun. Math.
  Phys.} {\bf 85} (1982) 481--490.

\bibitem{Gaiotto:2015zta}
D.~Gaiotto and A.~Kapustin, ``{Spin TQFTs and fermionic phases of matter},''
  {\em Int. J. Mod. Phys.} {\bf A31} (2016), no.~28n29 1645044,
  \href{https://arxiv.org/abs/1505.05856}{{\tt 1505.05856}}.

\bibitem{Chen:2017fvr}
Y.-A. Chen, A.~Kapustin, and {\DJ}.~Radi{\v c}evi{\'c}, ``{Exact bosonization
  in two spatial dimensions and a new class of lattice gauge theories},'' {\em
  Annals Phys.} {\bf 393} (2018) 234--253,
  \href{https://arxiv.org/abs/1711.00515}{{\tt 1711.00515}}.

\bibitem{Chen:2018nog}
Y.-A. Chen and A.~Kapustin, ``{Bosonization in three spatial dimensions and a
  2-form gauge theory},'' \href{https://arxiv.org/abs/1807.07081}{{\tt
  1807.07081}}.

\bibitem{Thorngren:2018bhj}
R.~Thorngren, ``{Anomalies and Bosonization},''
  \href{https://arxiv.org/abs/1810.04414}{{\tt 1810.04414}}.

\bibitem{Rothe:1992nt}
H.~J. Rothe, ``{Lattice gauge theories: An Introduction},'' {\em World Sci.
  Lect. Notes Phys.} {\bf 82} (2012). Chapters 7 and 8.

\bibitem{Chen:2011pg}
X.~Chen, Z.-C. Gu, Z.-X. Liu, and X.-G. Wen, ``{Symmetry protected topological
  orders and the group cohomology of their symmetry group},'' {\em Phys. Rev.}
  {\bf B87} (2013), no.~15 155114, \href{https://arxiv.org/abs/1106.4772}{{\tt
  1106.4772}}.

\bibitem{Kapustin:2014tfa}
A.~Kapustin, ``{Symmetry Protected Topological Phases, Anomalies, and
  Cobordisms: Beyond Group Cohomology},''
  \href{https://arxiv.org/abs/1403.1467}{{\tt 1403.1467}}.

\bibitem{Callan:1984sa}
C.~G. Callan, Jr. and J.~A. Harvey, ``{Anomalies and Fermion Zero Modes on
  Strings and Domain Walls},'' {\em Nucl. Phys.} {\bf B250} (1985) 427--436.

\bibitem{Komargodski:2017smk}
Z.~Komargodski, T.~Sulejmanpasic, and M.~{\"U}nsal, ``{Walls, anomalies, and
  deconfinement in quantum antiferromagnets},'' {\em Phys. Rev.} {\bf B97}
  (2018), no.~5 054418, \href{https://arxiv.org/abs/1706.05731}{{\tt
  1706.05731}}.

\end{thebibliography}\endgroup

\end{document}